# Press-Dyson Analysis of Asynchronous, Sequential Prisoner's Dilemma

Robert D. Young[*]

*Department of Physics, Illinois State University,
Normal, IL 61790-4560*

(Dated December 11, 2017)

Two-player games have had a long and fruitful history of applications stretching across the social, biological, and physical sciences. Most applications of two-player games assume synchronous decisions or moves even when the games are iterated. But different strategies may emerge as preferred when the decisions or moves are sequential, or the games are iterated. Zero-determinant strategies developed by Press and Dyson are a new class of strategies that have been developed for synchronous two-player games, most notably the iterated prisoner's dilemma. Here we apply the Press-Dyson analysis to sequential or asynchronous two-player games. We focus on the asynchronous prisoner's dilemma. As a first application of the Press-Dyson analysis of the asynchronous prisoner's dilemma, tit-for-tat is shown to be an efficient defense against extortionate zero-determinant strategies. Nice strategies like tit-for-tat are also shown to lead to Pareto optimal payoffs for both players in repeated prisoner's dilemma.

## 1. Introduction

The theory of games has its modern genesis in a classic book published in 1944 by von Neumann and Morgenstern [1]. Since then significant work by many researchers further developed the theory with applications to the social, behavioral, biological, and physical sciences. Prisoner's dilemma (PD) is a game developed in 1950 at RAND by Flood and Dresher [2]. Since its formal inception, PD has provided much insight and sometimes bafflement in understanding the resolution, and even intensification, of conflict and the development of cooperation in the real world. Axelrod describes many of the strategies used in attempts to solve the PD and reports on "tournaments" to test various strategies developed by himself and many others [3]. Sigmund has recently given a short, accessible book that treats PD as well as other classic games focusing on an evolutionary approach [4]. Recently, Cleveland, Liao, and Austin have applied a game theory approach, including PD, to the physics of cancer propagation [5]. Finally, the work of Brams [6] develops a dynamic theory of moves that includes PD and that has been applied to real world conflicts [7]. The wide scope and long history of game theory, prisoner's dilemma, and applications makes it surprising that a new technique and class of strategies for PD were recently discovered by Press and Dyson [8]. This discovery led to a resurgence of interest and new applications of the prisoner's dilemma, mostly in evolutionary biology [9]. The main purpose of this article is to extend the Press-Dyson analysis to a model of PD [10] that is sequential, asynchronous, and compatible with the theory of moves [6,7].

---

[*] Email: rdyoung@ilstu.edu

## II. Payoff Matrices for Prisoner's Dilemma

Prisoner's dilemma is developed around a landscape of payoffs for the players that is succinctly described in terms of an 2X2 matrix with ordered pair elements. This matrix captures both the decisions and the payoffs of the players [1-4,6]. Fig. 1 gives a standard form of the matrix.

$$\mathbf{X} \quad \begin{array}{c} \\ c \\ d \end{array} \begin{array}{cc} \mathbf{Y} \quad c & d \\ \begin{bmatrix} (R,R) & (S,T) \\ (T,S) & (P,P) \end{bmatrix} \end{array}$$

Figure 1. See text for discussion.

In Fig. 1, the two players are labeled X and Y. Player X is blue, and Y is red. Strategies of play for the players are labeled by c (cooperate) and d (defect). The verb cooperate means a move that the moving player judges to lead to a positive result for both players while defect means a move that the moving player judges to lead to a positive result for the moving player but a negative result for the other player. The ordered pairs in the four quadrants enclosed by brackets are payoffs of the two players based on the externality of interest. This can be a preference rank or a payoff. The language of payoffs will be used below. The first entry of the ordered pair corresponds to the payoff for player X, the second entry to the payoff for player Y. If each player cooperates the payoff is $R$ to each. If each player defects the payoff is $P$ to each. If one player cooperates and the other player defects, then the player that cooperates gets payoff $S$ and the player that defects get payoff $T$. The classic PD puts the following two conditions on the four payoffs: $T > R > P > S$ and $2R > T + S$. The first condition ensures that mutual defection with payoff $(P,P)$ is a Nash equilibrium [1,4,11]. The second condition ensures that mutual-cooperation with payoff $(R,R)$ is the best outcome of the four ordered pairs, a Pareto optimal payoff for both players with the two payoff sets mentioned below [4,6]. There are no other conditions on the payoffs for the PD except that the payoffs are drawn from integers including zero. Axelrod [3] emphasized the payoff set $(T,R,P,S) = (5,3,1,0)$ while Brams emphasized $(T,R,P,S) = (4,3,2,1)$. Other asynchronous models place additional conditions (eg, $T + S = R + P$) on the payoff set and are not considered here [12]. The matrix in Fig. 1 includes both player payoffs (ordered pairs) and player moves (c or d). The move information is included twice, once by the color code and once by the c or d designation to the top and left of the array of ordered pairs. Analysis of the structure of the matrices for general 2X2 games can be quite complex and rich with interpretation [6,7,13,14]. Here we consider only the matrix in Fig. 1. The color code of payoffs is used to be clear regarding which payoffs are associated with which player. However, the payoffs themselves can be different for the different players.

### III. Asynchronous Prisoner's Dilemma

von Neumann and Morgenstern understood that some form of repeated play was necessary for game theory to correspond to real conflicts [1]. One attempt to address this issue was the iterated prisoner's dilemma (IPD) with repeated play of the PD and synchronous moves of players X and Y [3,4,8]. The assumptions of the IPD result in transition probability matrices for player moves that are Markov with row entries adding to unity. Markov chains then is the underlying mathematical theory for IPD [15,16]. This results in the need to define probabilities for moving for both players called player strategies based on the outcome of the previous game. Probabilities based only on memory of the preceding game (memory-1 probabilities) are sufficient to define the rules of play [8]. Press and Dyson developed an analytic technique based on the structure of the transition matrices and associated determinants for the IPD that proves one player (or even both), by selecting certain playing strategies, can establish a linear relation between the average payoffs for both players. This simple linear relation was new to game theory and led to much activity, for example [9,17,18]. Synchronous player moves have long been recognized as a limitation when applying IPD [6,7,10,12]. Here a version of an Asynchronous Prisoner's Dilemma (APD) due to Frean [10] is analyzed with the Press-Dyson analysis to show that a linear relation between average payoffs exist in the APD with properly selected playing strategies.

Fig. 2 shows the essential ideas of the APD.

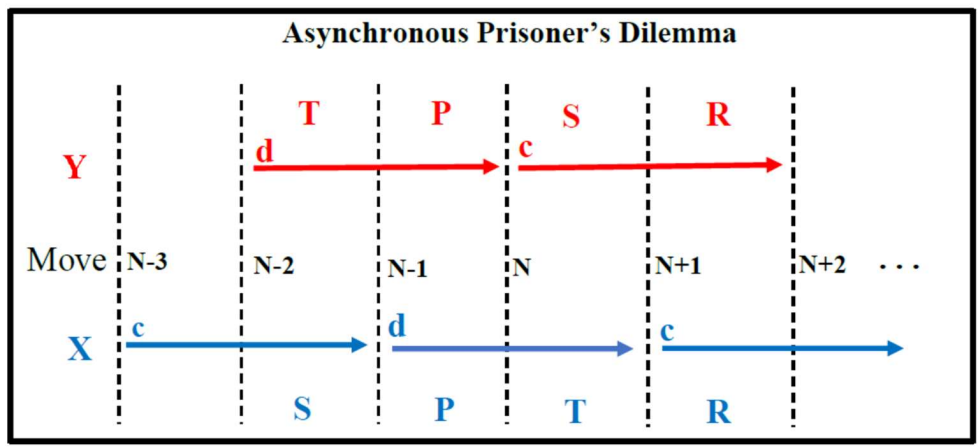

Figure 2. Asynchronous Prisoner's Dilemma (adapted from Frean [10]). X moves are on bottom; Y moves are on top. X moves at N-3, N-1, N+1 …; Y moves at N-2, N, N+2 …. Moves are asynchronous and sequential. Payoffs use T, R, P, S notation with X payoffs in blue on bottom and Y payoffs in red on top. Player moves are indicated by c and d in usual color and located near dotted line demarking moves. Players move with memory of the last move of the other player and that of themselves (memory-1 moves). Payoff to a player is determined by the player's last move and the other player's previous move. Specific moves and payoffs are examples. "Move" can also be interpreted as "time" [15,16].

From Fig. 2, player Y at move N cooperates (c) based on memory of X's defection (d) at move N-1 and Y's defection (d) at N-2. The resulting payoff for Y is S according to the payoff array in Fig. 1. At move N+1, player X cooperates (c) based on Y's cooperation at move N and X's defection at move N-1. The resulting payoff for X is T according to Fig. 1. In general, the 4-vector $\mathbf{p} = \left( p_1, p_2, p_3, p_4 \right)$ is the strategy of player X. $p_i$ are transition probabilities for player X to cooperate in the current game so $0 \le p_i \le 1$. The 4-vector $\mathbf{q} = \left( q_1, q_3, p_2, p_4 \right)$ is the strategy of player Y. $q_i$ are also transition probabilities for player Y to cooperate in the current game so $0 \le q_i \le 1$. The Markov transition matrix $\mathbf{M}_X$ for player X moving along the bottom of Fig. 2 is given by Frean [10] and reproduced in the Appendix, Eq. (A1), with further explanation. The Markov transition matrix $\mathbf{M}_Y$ for player Y moving along the top of Fig. 2 is also in the Appendix, Eq. (A2). The Markov matrix $\mathbf{M}_1 = \mathbf{M}_X \mathbf{M}_Y$ in the Appendix, Eq. (A3), is the transition probability matrix for the asynchronous, sequential game involving two players with player X moving first [10,12]. The Markov matrix $\mathbf{M}_2 = \mathbf{M}_Y \mathbf{M}_X$ in the Appendix, Eq. (A4), is the transition probability matrix for the asynchronous, sequential game with player Y moving first [10]. The following analysis of these transition probability matrices for the APD in Fig. 2 follows Press and Dyson [8]. The form of transition matrices do not depend on the exact payoffs so do not depend on the conditions of the PD.

Both $\mathbf{M}_1$ and $\mathbf{M}_2$ have a unit eigenvalue. The matrices $\bar{\mathbf{M}}_i \equiv \mathbf{M}_i - \mathbf{I}$ are both singular with zero determinants and stationary row eigenvectors $\mathbf{v}_i$ so $\mathbf{v}_i \bar{\mathbf{M}}_i = \mathbf{0}$ for $i = 1, 2$. Cramer's rule then implies that $\text{Adj}\left( \bar{\mathbf{M}}_i \right) \bar{\mathbf{M}}_i = \det\left( \bar{\mathbf{M}}_i \right) \mathbf{I} = \mathbf{0}$ where Adj refers to the adjugate matrix. The average of an arbitrary vector $\mathbf{f}$ given by the transpose $\mathbf{f}^{\mathrm{T}} = \left( f_1, f_2, f_3, f_4 \right)$ is expressed in terms of the stationary (probability) vectors by

$$\left\langle \mathbf{f} \right\rangle_i = \mathbf{v}_i \cdot \mathbf{f} = D_i \left( \mathbf{p}, \mathbf{q}, \mathbf{f} \right). \tag{1}$$

The determinants $D_i$ are given by

$$D_1 \left( \mathbf{p}, \mathbf{q}, \mathbf{f} \right) = \det \begin{vmatrix} -1 + p_1 q_1 & -1 + p_1 & \left( 1 - p_1 \right) q_2 & f_1 \\ p_2 q_3 & -1 + p_2 & \left( 1 - p_2 \right) q_4 & f_2 \\ p_3 q_1 & p_3 & -1 + \left( 1 - p_3 \right) q_2 & f_3 \\ p_4 q_3 & p_4 & \left( 1 - p_4 \right) q_4 & f_4 \end{vmatrix}, \tag{2}$$

and

$$D_2 \left( \mathbf{p}, \mathbf{q}, \mathbf{f} \right) = \det \begin{vmatrix} -1 + p_1 q_1 & p_2 \left( 1 - q_1 \right) & -1 + q_1 & f_1 \\ p_1 q_3 & -1 + p_2 \left( 1 - q_3 \right) & q_3 & f_2 \\ p_3 q_2 & p_4 \left( 1 - q_2 \right) & -1 + q_2 & f_3 \\ p_3 q_4 & p_4 \left( 1 - q_4 \right) & q_4 & f_4 \end{vmatrix}. \tag{3}$$

The two determinants $D_i(\mathbf{p}, \mathbf{q}, \mathbf{f})$ are each obtained from the matrices $\bar{\mathbf{M}}_i$ with the fourth column removed and replaced by $\mathbf{f}$. Furthermore, $D_1(\mathbf{p}, \mathbf{q}, \mathbf{f})$ has the first column of $\bar{\mathbf{M}}_1$ added to its second column, and $D_2(\mathbf{p}, \mathbf{q}, \mathbf{f})$ has the first column of $\bar{\mathbf{M}}_2$ added to its third column. Inspection of the determinants $D_i(\mathbf{p}, \mathbf{q}, \mathbf{f})$ shows that each has one column that can be under the control of one player. This means that zero-determinant strategies (ZDS) are possible [8]. A difference between this APD and the IPD of Press and Dyson is that Eqs. (2) and (3) show that only player X can implement a ZDS strategy when X moves first, and only player Y when Y moves first.

The payoff vector for player X is $\mathbf{S}_X = (R, S, T, P)$ and that for player Y is $\mathbf{S}_Y = (R, T, S, P)$. The average payoffs are

$$\left\langle S_k \right\rangle_i = \frac{\mathbf{v}_i \cdot \mathbf{S}_k^{\mathrm{T}}}{\mathbf{v}_i \cdot \mathbf{1}} = \frac{D_i(\mathbf{p}, \mathbf{q}, \mathbf{S}_k^{\mathrm{T}})}{D_i(\mathbf{p}, \mathbf{q}, \mathbf{1})} \tag{4}$$

where T denotes transpose, $i = 1, 2$ and $k = X, Y$. There are then four average payoffs, two for each player and two for each order of play. A Markov chain implies that the number of moves is large. In the resulting "fog of play", knowledge of which player moved first is obscured or lost. This means that the final payoffs for each player are obtained by equal weighted averages [10],

$$S_k = \frac{1}{2}\left(\left\langle S_k \right\rangle_1 + \left\langle S_k \right\rangle_2\right), \tag{5}$$

$k = X, Y$. To implement ZDS when player X moves first, X choses a strategy

$$\tilde{\mathbf{p}} = \left(-1 + p_1, -1 + p_2, p_3, p_4\right) = \alpha \mathbf{S}_X + \beta \mathbf{S}_Y + \gamma \mathbf{1} \tag{6}$$

and ensures that the average payoffs of the two players are linearly related,

$$\alpha \left\langle S_X \right\rangle_1 + \beta \left\langle S_Y \right\rangle_1 + \gamma = 0. \tag{7}$$

It should be noted that $\mathbf{p} = (p_1, p_2, p_3, p_4)$ is the actual strategy given by the value of $\tilde{\mathbf{p}}$. When player Y moves first, Y choses a strategy from the condition,

$$\tilde{\mathbf{q}} = \left(-1 + q_1, q_3, -1 + q_2, p_4\right) = \alpha' \mathbf{S}_X + \beta' \mathbf{S}_Y + \gamma' \mathbf{1} \tag{8}$$

and ensures that the average payoffs of the two players a linearly related,

$$\alpha' \left\langle S_X \right\rangle_2 + \beta' \left\langle S_Y \right\rangle_2 + \gamma' = 0. \tag{9}$$

This analysis shows that the essential idea of Press-Dyson analysis for deriving ZDS strategies holds in the APD.

## IV.  Application to Central Role of Nice Strategies including TFT

Axelrod established that the famous tit-for-tat (TFT) strategy submitted by Rapoport was efficient since TFT "won" both tournaments [3].  As an example of the anlysis developed here, we see how efficient TFT is against an extortionate ZDS.  When X moves first, an extortionate ZDS is defined by X selecting the strategy determined by [8],

$$\tilde{\mathbf{p}} = \left( -1 + p_1, -1 + p_2, p_3, p_4 \right) = \phi \left[ \left( \mathbf{S}_X - P\mathbf{1} \right) - \chi \left( \mathbf{S}_Y - P\mathbf{1} \right) \right] . \qquad (10)$$

This choice of strategy for X means that the parameters in Eq. (6) are re-expressed as $\alpha = \phi$, $\beta = \chi\phi$, and $\gamma = \phi\left(\chi - 1\right)P$.  Assuming the payoff $P$ is otherwise known, the extortionate ZDS redefines the parameters so that now there are only two free parameters $\chi$ and $\phi$, since $\gamma = \left(\beta - \alpha\right)P$.  It is clear also that the extortionate ZDS requires that $p_4 = 0$ so that X never cooperates once both players defect (dd).  Press and Dyson work out the details for the other three probabilities.  We do not need that information when Y plays the TFT strategy, $q_1 = q_3 = 1$ and $q_2 = q_4 = 0$.  To see this, Eqs. (2) and (4) give $\left\langle \mathbf{S}_X \right\rangle_1 = P$ and $\left\langle \mathbf{S}_Y \right\rangle_1 = P$.  Furthermore, Eqs. (3) and (4) give $\left\langle \mathbf{S}_X \right\rangle_2 = P$ and $\left\langle \mathbf{S}_Y \right\rangle_2 = P$ using the same strategies for X and Y.  With Eq. (5) this means that the average payoffs for X and Y are equal, $S_X = S_Y = P$, which is the Nash equilibrium for the PD.  Thus, even when X plays an extortionate ZDS, Y can always ensure a "fair" payoff equal to that of X, even if the payoffs are the Nash equilibrium with payoffs less than $T$ or $R$.  This is true independent of the exact extortionate strategy of X.   The TFT strategy is a ZDS strategy for X with parameters $\chi = 1$ and $\phi = 1/5$.  Without necessarily being informed of general ZDS analysis, however, Y has chosen a simple strategy that ensures a fair payoff.

The key to the successful defense of Y against extortion by X is the selection of the probability $q_4 = 0$ so that Y never cooperates once both players defect (dd).  This is clear from inspection of the determinants in Eqs. (2) and (3) where the conditions $p_4 = q_4 = 0$ means that all determinants derived from Eqs. (2) and (3) are proportional to $f_4$ and the resulting 3X3 minors obtained by eliminating the fourth row and column are all equal (and non-zero).  This is the reason $\left\langle S_k \right\rangle_i = P$ for all $k$ and $i$.

There clearly is more to TFT since ensuring a $\left(P, P\right)$ against extortionate strategies wouldn't be enough to win the Axelrod tournaments.  Insight is gained by inspecting Eqs. (2) and (3) and asking what is the simplest set of strategies to reach $\left(R, R\right)$, the mutual payoff that is Pareto optimal, $2R > T + S$?  This can be accomplished by having both determinants in Eqs. (2) and (3) proportional to $f_1$ which equals $R$ when $\mathbf{f} = \mathbf{S_X}, \mathbf{S_Y}$. A class of cooperative strategies has $p_1 = q_1 = 1$ so that both players always cooperate when they have cooperated on their previous moves. This condition ensures a payoff $R$ against many cooperative strategies. To see this, using Eq. (B2) in (4) gives $\langle \mathbf{S_X} \rangle_1 = \langle \mathbf{S_Y} \rangle_1 = R$ for X moving first. Using Eq. (B5) in (4) gives $\langle \mathbf{S_X} \rangle_2 = \langle \mathbf{S_Y} \rangle_2 = R$ for Y moving first. This means that the average payoffs are given by $S_X = S_Y = R$ so that the APD results in the Pareto optimal payoff $(R, R)$ when only cooperative strategies with $p_1 = q_1 = 1$ are chosen. Although $R$ is less than $T$, players will avoid the sucker payoff $S$ so $T$ payoffs will not happen with significant frequency in a repeated PD. The determinants $\bar{D}_i (\mathbf{p}, \mathbf{q}) \big|_{p_1 = q_1 = 1}$ in Eqs. (B3) and (B6) allow for a large (infinite) number of cooperative strategies providing payoffs of $R$. One such strategy of course is TFT. This analysis suggests why so called "nice" strategies scored well in Axelrod's tournaments [3]. Press and Dyson also pointed out that TFT, played by X will result in the payoff $(R, R)$ when Y is cooperative, $\mathbf{q} = (1,1,1,1)$, a clearly nice strategy.

The analysis of the last paragraph did not use the methodology of ZDS. Press-Dyson analysis can be used without identifying ZDS strategies however as demonstrated above. If ZDS analysis were used, then the strategy for X can be selected by the condition

$$\tilde{\mathbf{p}} = \left( -1 + p_1, -1 + p_2, p_3, p_4 \right) = \phi \left[ \left( \mathbf{S_X} - R\mathbf{1} \right) - \chi \left( \mathbf{S_Y} - R\mathbf{1} \right) \right] \qquad (11)$$

which results in the probability $p_1 = 1$. Notice the substitution of $R$ rather than $P$ in Eq. (11) when compared to Eq. (10). The same line of analysis as in Press and Dyson can then be carried out with Y having strategy $\mathbf{q} = (1, q_3, q_2, q_4)$ rather than a full cooperative strategy. The results of course would be the same as in the last paragraph. The point here is that ZDS does not automatically imply extortion or forcing, but can equally be used to analyze the implications of cooperation. Finally, the case of both X and Y using TFT strategies is a singular case. In the fog of play, the payoffs depend on what moves were made after the Markov chain analysis was implemented. Eventually chains of cooperation are developed with payoffs $(R, R)$, or chains of defection are developed with payoffs $(P, P)$. The conclusions of this section also follow from the synchronous PD of Press and Dyson (see their Fig. 2B [8]) and are consequences of repeated PD not of asynchronous or synchronous moves.

## V. Conclusions

The asynchronous or sequential prisoner's dilemma is a robust model for conflict situations in which players alternate moves, and payoffs are determined by the current player's move and the move of the previous player. This asynchronous prisoner's dilemma places no restrictions on the values of the payoff set R, S, T, P other than the conditions determining a prisoner's dilemma. In fact, the two conditions determining the prisoner's dilemma can be removed since the analysis is not dependent on these conditions. Two players can also have different payoffs. This analysis can be applied to many 2X2 games [13,14]. The Press-Dyson analysis resulting in zero-determinant strategies is successfully applied to the asynchronous prisoner's dilemma. As a first example, TFT is shown to be an efficient defense against extortionate zero-determinant strategies by guaranteeing a fair result, namely the Nash equilibrium. In addition, the minimally cooperative strategy of both players cooperating when they cooperated on the previous move is also shown to be sufficient to reach the Pareto optimal $(R, R)$ payoff in the prisoner's dilemma. This suggests why nice strategies fare well in Axelrod tournaments.

<div align="center">

**Acknowledgement**

</div>

I thank Steve Walker for many stimulating discussions during runs and walks in the forest.

<div align="center">

**Appendix A**

</div>

The Markov transition matrix $\mathbf{M}_X$ for player X moving as in Fig. 2 is given by [10],

$$\mathbf{M}_X = \begin{array}{c} \\ cc \\ cd \\ dc \\ dd \end{array} \begin{array}{cccc} cc & cd & dc & dd \\ \begin{pmatrix} p_1 & 0 & 1-p_1 & 0 \\ 0 & p_2 & 0 & 1-p_2 \\ p_3 & 0 & 1-p_3 & 0 \\ 0 & p_4 & 0 & 1-p_4 \end{pmatrix} \end{array} . \tag{A1}$$

The notation, cc, cd, dc, and dd, above and to the left of the transition probability matrix correspond to the moves of player X and Y according to the ordering, xy. The prior move is given down the left and the current move is given along top. This sequence of moves corresponds to the chain of moves in Fig. 2 where X only moves. The blue transition probabilities $p_i$ in Eq. (A1) are conditional probabilities for X to cooperate (c) in the current round depending on the state of the previous round: $p_1 = \text{Prob}(X \rightarrow c \mid cc)$; $p_2 = \text{Prob}(X \rightarrow c \mid cd)$; $p_3 = \text{Prob}(X \rightarrow c \mid dc)$; $p_4 = \text{Prob}(X \rightarrow c \mid dd)$. It follows that all entries in which Y moves must vanish, corresponding to the red 0 in $\mathbf{M}_X$. The Markov transition matrix $\mathbf{M}_Y$ for player Y moving as in Fig. 2 is

$$\mathbf{M}_Y = \begin{pmatrix} q_1 & 1-q_1 & 0 & 0 \\ q_3 & 1-q_3 & 0 & 0 \\ 0 & 0 & q_2 & 1-q_2 \\ 0 & 0 & q_4 & 1-q_4 \end{pmatrix} . \tag{A2}$$

The strategy vector for player X is $\mathbf{p} = (p_1, p_2, p_3, p_4)$, and that for player Y is $\mathbf{q} = (q_1, q_3, q_2, q_4)$ [10].

The Markov matrix $\mathbf{M}_1 = \mathbf{M}_X \mathbf{M}_Y$ is the transition matrix for the game of alternating, sequential game with player X moving first [10, 12],

$$\mathbf{M}_1 = \begin{pmatrix} p_1 q_1 & p_1(1-q_1) & (1-p_1)q_2 & (1-p_1)(1-q_2) \\ p_2 q_3 & p_2(1-q_3) & (1-p_2)q_4 & (1-p_2)(1-q_4) \\ p_3 q_1 & p_3(1-q_1) & (1-p_3)q_2 & (1-p_3)(1-q_2) \\ p_4 q_3 & p_4(1-q_3) & (1-p_4)q_4 & (1-p_4)(1-q_4) \end{pmatrix}. \tag{A3}$$

The Markov matrix $\mathbf{M}_2 = \mathbf{M}_Y \mathbf{M}_X$ is the transition matrix for the alternating, sequential game with player Y moving first [10],

$$\mathbf{M}_2 = \begin{pmatrix} p_1 q_1 & p_2(1-q_1) & (1-p_1)q_1 & (1-p_2)(1-q_1) \\ p_1 q_3 & p_2(1-q_3) & (1-p_1)q_3 & (1-p_2)(1-q_3) \\ p_3 q_2 & p_4(1-q_2) & (1-p_3)q_2 & (1-p_4)(1-q_2) \\ p_3 q_4 & p_4(1-q_4) & (1-p_3)q_4 & (1-p_4)(1-q_4) \end{pmatrix}. \tag{A4}$$

## Appendix B

From Eq. (2),

$$D_1(\mathbf{p}, \mathbf{q}, \mathbf{f})\big|_{p_1 = q_1 = 1} = \det \begin{vmatrix} 0 & 0 & 0 & f_1 \\ p_2 q_3 & -1 + p_2 & (1-p_2)q_4 & f_2 \\ p_3 & p_3 & -1 + (1-p_3)q_2 & f_3 \\ p_4 q_3 & p_4 & (1-p_4)q_4 & f_4 \end{vmatrix}, \tag{B1}$$

so that

$$D_1(\mathbf{p}, \mathbf{q}, \mathbf{f})\big|_{p_1 = q_1 = 1} = -f_1 \, \bar{D}_1(\mathbf{p}, \mathbf{q})\big|_{p_1 = q_1 = 1} \tag{B2}$$

where

$$\bar{D}_1(\mathbf{p}, \mathbf{q})\big|_{p_1 = q_1 = 1} = \det \begin{vmatrix} p_2 q_3 & -1 + p_2 & (1-p_2)q_4 \\ p_3 & p_3 & -1 + (1-p_3)q_2 \\ p_4 q_3 & p_4 & (1-p_4)q_4 \end{vmatrix} \neq 0. \tag{B3}$$

From Eq. (3)

$$D_2\left(\mathbf{p},\mathbf{q},\mathbf{f}\right)\Big|_{p_1=q_1=1} = \det\begin{vmatrix} 0 & 0 & 0 & f_1 \\ q_3 & -1+p_2\left(1-q_3\right) & q_3 & f_2 \\ p_3q_2 & p_4\left(1-q_2\right) & -1+q_2 & f_3 \\ p_3q_4 & p_4\left(1-q_4\right) & q_4 & f_4 \end{vmatrix}, \tag{B4}$$

so that

$$D_2\left(\mathbf{p},\mathbf{q},\mathbf{f}\right)\Big|_{p_1=q_1=1} = -f_1\,\overline{D}_2\left(\mathbf{p},\mathbf{q}\right)\Big|_{p_1=q_1=1} \tag{B5}$$

where

$$\overline{D}_2\left(\mathbf{p},\mathbf{q}\right)\Big|_{p_1=q_1=1} = \det\begin{vmatrix} q_3 & -1+p_2\left(1-q_3\right) & q_3 \\ p_3q_2 & p_4\left(1-q_2\right) & -1+q_2 \\ p_3q_4 & p_4\left(1-q_4\right) & q_4 \end{vmatrix} \neq 0. \tag{B6}$$